\documentstyle[12pt]{article}

\global\arraycolsep=1pt
\begin{document}

\hfill    SISSA/ISAS 43/93/EP

\hfill    June 1993

\begin{center}
\vspace{24pt}
{\large \bf COVARIANT PAULI-VILLARS REGULARIZATION\\
OF QUANTUM GRAVITY AT THE ONE LOOP ORDER}
\vspace{24pt}

{\sl Damiano Anselmi}

\vspace{6pt}

SISSA - International School for Advanced Studies, via Beirut 2-4,
I-34100 Trieste, Italy\\
and I.N.F.N. - Sezione di Trieste, Trieste, Italy\\

\vspace{12pt}

\end{center}

\vspace{24pt}

\begin{center}
{\bf Abstract}
\end{center}

\vspace{12pt}

\noindent
We study a regularization of the Pauli-Villars kind
of the one loop gravitational divergences in any dimension.
The Pauli-Villars fields are massive particles coupled
to gravity in a covariant and nonminimal way, namely one real tensor
and one complex vector.
The gauge is fixed by means of the unusual gauge-fixing that
gives the same effective action as in the context of
the background field method. Indeed, with
the background field method it is simple to see that the
regularization effectively works.
On the other hand, we show that in the usual formalism (non background)
the regularization cannot work with each gauge-fixing.
In particular, it does not work with the usual one.
Moreover, we show that, under a suitable choice of the
Pauli-Villars coefficients,
the terms divergent in the Pauli-Villars
masses can be corrected by the Pauli-Villars
fields themselves. In dimension four, there is no need to add counterterms
quadratic in the curvature tensor to the Einstein action
(which would be equivalent to the introduction of new coupling constants).
The technique also works when matter is coupled to gravity.
We discuss the possible consequences of this approach, in particular the
renormalization of Newton's coupling constant and the appearance of two
parameters in the effective action, that seem to have physical
implications.
Whether our procedure can be extended to all loops or can make
quantum gravity a finite theory
are matters that deserve further investigations.
\vfill
\eject
\section{Introduction}
\label{intro}

In a couple of recent papers \cite{anselmi,anselmi2}, we began a program
intended to
formulate and study a regularization of quantum gravity at the one loop order
that preserves explicit BRS invariance and that
is definable at the level of the functional integral (to be compared,
for example, with dimensional
regularization, which is definable only at the level of Feynmann diagrams).

We first discussed the problem of the
functional measure.
It is related to the maximal one loop divergences \cite{anselmi}.
This is a clear
example of something that is invisible as far as dimensional
regularization is concerned and
that gives some insight into the program of correctly defining
the functional integral. We find that,
using a naive cut-off, the maximal divergences (which are not BRS invariant)
are not cancelled
by the BRS invariant measure \cite{anselmi,anselmi2}.
Consequently, in order to avoid the explicit breaking
of BRS invariance that occurs when employing a naive regularization technique
(save particular cases, for instance when the theory is supersymmetric
\cite{anselmi2}), one needs a suitable non-BRS-invariant counterterm
or, equivalently, a non-BRS-invariant measure.
We thus focused our
attention on the Pauli-Villars regularization,
in particular on the technique that works at the
one loop order (sometimes known as Pauli-Bethe). Indeed, it is possible to
give a formulation (called by us the DTNP formalism
\cite{olandesi}) that explicitly preserves covariance:
both the regularized lagrangian and the measure and the regularization
procedure turn out to be BRS-invariant. This
is very convenient, since we wish to characterize the one loop divergences
in a simple way. Moreover, the Pauli-Villars regularization possesses
some properties that are not common to the other regularization techniques. In
particular, since there is more that one regulator,
one wonders if it is possible
to make the Pauli-Villars (PV) regulators cancel the
divergences that they themselves produce. In Ref.\
\cite{anselmi} this was done  for the maximal divergences:
on the one hand, the Pauli-Villars regularization
cancels the maximal divergences that would appear with a naive cut-off and
returns divergences in the masses of the PV fields; on the other hand,
the maximal divergences
in the PV masses
can be made disappear by adding a condition on the PV coefficients, with
no explicit introduction of counterterms (see formula (2.7) of Ref.\
\cite{anselmi}).

In this paper we make the same thing
for the nonmaximal one loop divergences and explore the possible consequences.
Our aim is to formulate a one loop regularization of the Pauli-Villars
kind that does
not break explicit BRS invariance. We call this technique
{\sl covariant Pauli-Villars regularization}
of quantum gravity at the one loop order.
After doing this, we discuss its implications. In particular, the
properties of the Pauli-Villars regularization permit to avoid any
modification
of the starting lagrangian, such as the introduction of new
coupling constants.
This is interesting, because
in dimension four
it saves the addition of counterterms quadratic in the Riemann tensor.
If one adds these counterterms to the starting lagrangian,
the resulting theory is no longer Einstein gravity,
i.e.\ it is no longer the theory one started from. Moreover,
higher derivative quantum gravity is renormalizable,
but not unitary \cite{stelle}.
In the context of our analysis the finiteness of pure
gravity in dimension four
\cite{thooft} is nothing more than an accident. Our reasonings
apply in general, that is in any dimension (greater that two)
and in presence of matter.

Let us briefly mention how the various kinds of divergences
are treated within our
regularization procedure.

i) The maximal divergences in the effective
action have the form of a cosmological term.
They can be made disappear in the way that we recalled
a moment ago, i.e.\ by imposing a
condition on the PV coefficients.
If our starting theory
itself contained a cosmological term, there would be no need
to make the maximal divergences
vanish in this way, because they could be cancelled by renormalizing
the cosmological constant.

ii) The logarithmic divergences of four dimensional quantum gravity are
quadratic in the Riemann tensor. They can be cancelled by imposing a further
condition on the PV coefficients. This seems to be the only way to get
rid of them without modifying the starting lagrangian. The price one
has to pay is the appearance of two arbitrary constants in
the effective action.
They should be fixed experimentally and seem to be something like
``quantum coupling constants",
i.e.\ coupling constants that are present in the
effective action, although they are absent in the classical action.

iii) In dimension four, there are not only maximal divergences and logarithmic
divergences,
but also quadratic divergences. The terms of the one loop effective action that
diverge quadratically in the PV masses have the same form as the terms of the
starting lagrangian (the Einstein lagrangian plus the matter lagrangian).
Thus, the quadratic divergences are responsible for the renormalization
of the Newton
coupling constant and the wave function renormalization.

Summasizing, our technique differs from the usual renormalization procedure
in the sense that not all (one loop) divergences are absorbed by a
renormalization of physical parameters, but some kinds of divergences
are made to vanish by extra conditions on the PV coefficients.
In particular, in four dimensional quantum gravity, logarithmic divergences
(and eventually quartic divergences, if one wants no cosmological term)
are avoided by imposing suitable extra relations between the PV masses
and the PV coefficients, while quadratic divergences can be absorbed by a
renormalization of the Newton coupling constant. Of course, quadratic
divergences can also be made to vanish by a further condition on PV
coefficients.
So far, we have not found an argument for privileging one of the two
ways of cancelling those divergent terms (like the quadratic divergences
of four dimensional quantum gravity) that have
the same form as the terms of the starting lagrangian.

In the case of an ordinary renormalizable quantum field theory, our technique
has no remarkable property, since there is no divergent term
that cannot be absorbed by a renormalization of a physical parameter
of the starting lagrangian. Along with our discussion, we shall have occasion
to explain why in the case of a renormalizable theory
our technique has the same effects as any ordinary one
(see section \ref{renorma}): one can consider it as
a manifestly gauge-invariant Pauli-Villars
regularization at the one loop
order, but nothing more.

The paper is organized as follows. In section \ref{2} we recall
the properties of the
DTNP formalism and give the conventions and the regularization conditions.
In section \ref{9} we show that a Pauli-Villars one loop regularization of
quantum gravity of the kind that we have in mind cannot work
with a generic gauge-fixing.
Indeed, it works with gauge-fixings that guarantee off-shell
covariance of the effective action, as in the case of the
background field method.
In section \ref{regula} we formulate the one loop regularization and
show that it effectively works, while in section \ref{renorma}
we discuss the properties
of our PV regularization, in particular what are the conditions
on the PV coefficients
that avoid the explicit introduction of counterterms. In section
\ref{coupling}
we analyse the renormalization of the Newton coupling constant
and the wave function renormalization. Moreover, we discuss the appearance of
two arbitrary parameters in the one loop effective action. Section \ref{concl}
contains our conclusions, while in the appendix the description of the
possible divergences appearing in the standard one loop computation
can be found.

\section{DTNP regularization formalism}
\label{2}

We employ the regularization formalism
described in Ref.\ \cite{olandesi},
which has the advantage of giving a
one loop regularization of the Pauli-Villars
kind while preserving gauge
invariance.
The DTNP formalism is based on a particular
definition of functional integration on
PV regulators. If $\chi_j$ is a PV bosonic field
and $A$ is a generic $\chi-$independent
infinite matrix, we define
\begin{equation}
\int {\cal D}^\prime \chi_j \hskip .2truecm
e^{\chi_j^T A \chi_j}=( {\rm det}
\hskip .1truecm A)^{c_j \over 2},
\label{dtnp}
\end{equation}
where $^T$ denotes transposition and $c_j$
is a coefficient associated to
the PV field $\chi_j$, as in ordinary
Pauli-Villars regularization.
The symbol ${\cal D}^\prime$ is to remember that,
even if Eq.\ (\ref{dtnp})
resembles the usual gaussian functional integral
(in which, however,
$c_j$ should be $-1$), it is only a convenient
way of writing down an infinite
determinant.  Notice that the PV regulators have the same statistics
as the fields that they regulate.

The measure ${\cal D}^\prime \chi_j$
introduced in Eq. (\ref{dtnp})
is not invariant
under a shift of the integration variable.
However, within our formalism,
no contradiction can arise since we never need
to introduce PV sources
and a shift of the integration variables is never
necessary to our reasonings.
The rule for a
homogeneous linear
change of variables is
\begin{equation}
{\cal D}^\prime \tilde\chi_j=( {\rm det}
\hskip .2truecm K)^{-c_j}
{\cal D}^\prime \chi_j,
\label{det}
\end{equation}
where $\tilde\chi_j=K\chi_j$ \cite{olandesi,anselmi}.

Now let us give the other regularization
conventions.

For PV boson fields $\chi_j$
with masses $M_j$ we impose the following regularization
conditions
\begin{eqnarray}
\sum_{j=1}^n c_j &=&c,
\nonumber\\
\sum_{j=1}^n c_j \hskip .1truecm  (M_j^2)^p&=&0
\hskip 1truecm
0<p\leq\left[{r\over 2}\right],
\label{pvcond}
\end{eqnarray}
where $p$ is integer and $\left[r\over 2\right]$ is
the integral part of
$r\over 2$. Here $r$ is the space-time dimension.
In Ref.\ \cite{anselmi} $c$ was generic (it was fixed by requiring
the regularization of maximal divergences). In the present paper
we shall need only the cases $c=1$ and $c=-2$.

In Ref.\ \cite{anselmi} [formula (2.1)]
we added a condition on the coefficients $c_j$
so as to renormalize the maximal divergences, namely
\begin{eqnarray}
\sum_{j=1}^n \hskip .1truecm  c_j \hskip .1truecm
M_j^r \hskip .1truecm
{\rm ln} \hskip .05truecm
\left({M_j^2 \over \mu^2}\right)=0 &\quad\quad
{\rm if\hskip .2truecm} r {\rm \hskip .2truecm
is\hskip .2truecm even,}
\nonumber\\
\sum_{j=1}^n \hskip .1truecm  c_j \hskip .1truecm
M_j^r=0 &\quad\quad
{\rm if\hskip .2truecm} r {\rm \hskip .2truecm
is\hskip .2truecm odd,}
\label{logcond}
\end{eqnarray}
where $\mu$ is a certain scale that makes the
argument of the logarithm
dimensionless\footnotemark\footnotetext{Conditions similar to the first
of (\ref{logcond}) (with $M_j^r$ replaced by $M_j^2$)
can also be found in the article by W.\ Pauli and
F.\ Villars \cite{paulivillars}. They were studied for the
calculation of diagrams at zero external momentum in Q.E.D.
}. Condition (\ref{logcond})
does not depend on $\mu$ by
virtue of the fact that
according to Eq.\ (\ref{pvcond})
$\sum_{j=1}^n \hskip .1truecm  c_j \hskip .1truecm
M_j^r=0$ in even dimensions.
Condition (\ref{logcond}) is very important for the reasonings of Ref.s
\cite{anselmi,anselmi2}. Indeed, after regularizing the maximal divergences
that appear with the naive cut-off, it permits to renormalize the maximal
divergences in the PV masses.
In the present investigation, we shall see that it is possible to introduce
other conditions similar to (\ref{logcond}) (see section \ref{renorma})
that perform the renormalization
of the other one loop divergences (the nonmaximal ones),
without the explicit introduction of
those counterterms that would modify the starting lagrangian.
In particular, this avoids the problem of counterterms quadratic in the
Riemann tensor in four dimensions,
which would imply the introduction of new coupling constants.

The number $n$ of PV fields $\chi_j$ must be
sufficient to assure the existence of a simultaneous solution
of equations (\ref{pvcond})
and (\ref{logcond}) and the further conditions that we shall impose
in section \ref{renorma}. It will be shown that $n=2r$ is sufficient to our
purposes.

\section{Preferred gauge-fixings}
\label{9}

The question is if the regularization scheme we employed in Ref.s
\cite{anselmi,anselmi2} can
regularize the nonmaximal divergences at the one loop order.
The maximal divergences ($\delta^{(r)}(0)$) correspond
to $\Lambda^r$, where $\Lambda$ is a naive cut-off.
For $r=4$ there are also divergences of the kind
$\Lambda^2$ and ${\rm ln}
\Lambda^2$. For $r>4$ there are further
divergences, but in this section
we shall restrict ourselves to the case $r=4$. We show that our regularization
technique cannot work with a generic gauge-fixing. The only possibility is
to make it work with a preferred kind of gauge-fixing.

Let us make an introductory remark.
There is the possibility
of introducing lots of nonminimal couplings in the lagrangians of the
PV regulators (these lagrangians can be found in Ref.\ \cite{anselmi}),
that do not
modify the maximal divergences, nor they affect the free lagrangian.
For instance, take a PV vector $W_\mu$. We can add
\begin{equation}
R\; W_\mu W^\mu\sqrt{-g}, \hskip .1truecm \hskip .1truecm
{\rm or} \hskip .1truecm \hskip .1truecm R^{\mu \nu}\;
W_\mu W_\nu
\sqrt{-g}.
\label{nonmin}
\end{equation}
For a PV tensor $T_{\mu\nu}$, there are the following five possibilities
\begin{eqnarray}
R\; T_{\mu \nu}T^{\mu \nu}\sqrt{-g}, \hskip .1truecm\hskip .1truecm
\hskip .1truecm\hskip .1truecm
R\; T^2\sqrt{-g}, \hskip .1truecm\hskip .1truecm\hskip .1truecm
\hskip .1truecm R^{\mu \nu}\; T_{\mu \nu}T\sqrt{-g},
\nonumber\\
R^{\mu \nu}\; T_{\mu \rho}{T^\rho}_\nu\sqrt{-g}, \hskip .1truecm
\hskip .1truecm\hskip .1truecm\hskip .1truecm\hskip .1truecm\hskip .1truecm
R^{\mu \nu, \rho\sigma}\; T_{\mu \rho}T_{\nu \sigma}\sqrt{-g},
\label{nonmin2}
\end{eqnarray}
where $T=T_{\mu\nu}g^{\mu\nu}$.
The constants in front of these couplings are numbers that should
be determined by requiring the regularization of nonmaximal divergences,
to mimic what we did for the measure and the PV coefficient $c$ in Ref.\
\cite{anselmi}. They were indeed fixed by requiring the regularization
of the maximal divergences. In the present case,
the number of unknowns is greater than the number of conditions
that should be
satisfied
(remember we can make a convex linear combination of regulators
without affecting the maximal divergences \cite{anselmi}).
Nevertheless, we now show that the requirement of regularization
cannot be satisfied with the usual gauge-fixing.
In particular, the logarithmic divergences cannot be regularized by the PV
fields, in the context of the approach that we have in mind (that aims to
preserve explicit BRS invariance).
In the following reasoning, we shall use the
correspondence between logarithmic divergences and the
poles that one finds in dimensional regularization (${2\over \varepsilon}=
\ln {\Lambda^2\over \mu^2}$, where $\varepsilon=4-r$,
$\Lambda$ is the naive cut-off and $\mu$ is an arbitrary scale)
and the fact that
these divergences satisfy Ward identities.

We must give some notation (the same as in Ref.\ \cite{anselmi}),
to make the explanation clearer.
The gravitational variables we shall use in this section are
\begin{equation}
\tilde g^{\mu \nu} \equiv \sqrt{-g} g^{\mu \nu}
\equiv \eta^{\mu \nu}+
\kappa \phi^{\mu \nu},
\end{equation}
where $g^{\mu \nu}$ is the inverse of the metric tensor,
$\phi^{\mu \nu}$ is the quantum field and
$\eta^{\mu \nu}={ \rm diag}(1,-1,-1,\ldots)$;
$\kappa=\sqrt{32 \pi G},$ G being Newton's constant.
The indices of $\phi^{\mu \nu}$
will be lowered by means of $\eta_{\mu \nu}$.
We denote by $\tilde g_{\mu \nu}$ the inverse matrix
of $\tilde g^{\mu
\nu}$. The gravitational
lagrangian is \cite{medrano}
\begin{equation}
{\cal L}_{g}={1 \over 2
\kappa^2} \left( \tilde g^{\rho \sigma}
\tilde g_{\lambda \mu}
\tilde g_{k \nu}-{1 \over r-2} \tilde g^{\rho \sigma}
\tilde g_{\mu k} \tilde
g_{\lambda \nu}-2 \delta^{\sigma}_{k}
\delta^{\rho}_{\lambda}
\tilde g_{\mu \nu} \right)
{\tilde g}^{\mu k}_{\phantom{\mu k},\rho}
{\tilde g}^{\nu \lambda}_{\phantom{\nu\lambda},\sigma}
\label{lg}
\end{equation}
where a comma indicates ordinary
differentiation. $ {\cal L}_g$ is ${2\over
\kappa^2} \sqrt{-g}R$
up to a total derivative.

We choose the usual gauge-fixing term
\begin{equation}
 {\cal L}_{gf}=-{1 \over \alpha \kappa^2}
\partial_\mu \tilde
g^{\mu \nu}\partial_\rho
\tilde g^{\rho \nu^\prime}\eta_{\nu \nu^\prime}.
\label{gf}
\end{equation}
The ghost lagrangian corresponding to Eq.\
(\ref{gf}) is
\begin{eqnarray}
 {\cal L}_{gh}&=\bar C_\nu [ \partial^\rho
\partial_\rho
\delta_\mu^\nu-\kappa (
{\phi^{\rho \nu}}_{, \mu \rho}-\phi^{\rho \sigma}
\partial_\rho \partial_\sigma
\delta_\mu^\nu
-{\phi^{\rho \sigma}}_{, \rho} \partial_\sigma
\delta_\mu^\nu+
{\phi^\rho}_{\nu, \rho}
\partial_{\mu} ) ]  C^\mu.
\end{eqnarray}

We introduce sources $J^{\mu \nu}$
for $\phi_{\mu \nu}$ and sources $\xi^\mu$ for
$\bar C_\mu$. The sources for the
other fields are not necessary for our derivation.
Let us work in the context of dimensional regularization and without PV
fields, for now.
One can easily find, using BRS invariance, the identity
\begin{equation}
< J^{\mu \nu}\delta_{BRS}\phi_{\mu \nu}+\delta_{BRS}\bar C_\mu \xi^\mu
> =0.
\end{equation}
Differentiating in $J^{\mu \nu}$ and $\xi^\sigma$ and setting
$J^{\mu \nu}=0$ and $\xi^\mu=0$, we find
\begin{equation}
0=<\bar C_\rho (x)\delta_{BRS}\phi_{\mu\nu}(y)+
\delta_{BRS}\bar C_\rho(x)\phi_{\mu\nu}(y)>=0,
\end{equation}
that gives, in momentum space,
\begin{eqnarray}
{2 \over \alpha}&k^\rho<\phi_{\mu \nu}(k)\phi_{\rho \sigma}(-k)>
+(\eta_{\mu \nu}k_\rho-\eta_{\nu \rho}
k_\mu-\eta_{\mu \rho}k_\nu)< C^\rho (-k) \bar C_\sigma (k) >
\nonumber\\&
+\kappa k_\rho < C^\rho (-p) \phi_{\mu \nu}(p-k)\bar C_\sigma (k) >
-\kappa p^\rho < C_\mu (-p)\phi_{\nu \rho}(p-k) \bar C_\sigma (k) >
\nonumber\\&
-\kappa p^\rho < C_\nu (-p) \phi_{\mu \rho}(p-k) \bar C_\sigma (k)>=0,
\label{ward}
\end{eqnarray}
where convolution in momentum $p$ is understood (with measure
$d^4p\over (2 \pi)^4$)
when the momentum $p$ is repeated two times
in the arguments of the fields.
Contracting by $k_\mu$ and suitably manipulating, one can find the identity
\cite{medrano}
\begin{equation}
{2\over \alpha}k^\rho k^\mu < \phi_{\mu \nu}(k)\phi_{\rho \sigma}(-k)
>=-i \eta_{\nu \sigma}.
\end{equation}
In particular, the logarithmic divergences $\tilde\Pi_{\mu \nu,\rho \sigma}$
of the graviton self-energy $\Pi_{\mu\nu,\rho\sigma}$
(with propagators applied to the external legs)
satisfy
\begin{equation}
k_\rho k_\mu \tilde\Pi_{\mu \nu,\rho \sigma}=0.
\label{medrano}
\end{equation}
Following similar steps in the case of a simpler theory,
namely Yang-Mills theory, one finds
\begin{equation}
k_\mu k_\nu \tilde\Pi^{YM}_{\mu a,\nu b}=0.
\end{equation}
{}From this one immediately deduces
\begin{equation}
k_\nu \tilde\Pi^{YM}_{\mu a,\nu b}=0,
\end{equation}
by simply counting the Einstein indices at disposal:
indeed, the logarithmic divergences of the self-energy of a Yang-Mills
field are transverse.
Form (\ref{medrano}) one cannot deduce
\begin{equation}
k_\rho \tilde\Pi_{\mu \nu,\rho \sigma}=0,
\end{equation}
because there are more Einstein indices. Using the identity (\ref{ward})
we can easily find $k_\rho \tilde\Pi_{\mu\nu,\rho\sigma}$
by calculating only diagrams with ghosts, which are
considerably simpler. One finds
\begin{equation}
{2\over \alpha}k_\rho \tilde\Pi_{\mu \nu, \rho \sigma}=
-{k_\sigma \over 2}\left(\eta_{\mu \nu}-{k_\mu k_\nu \over k^2}\right)
{1\over 16\pi^2}\hskip .1truecm {\rm ln} \hskip .05truecm
{\Lambda^2\over \mu^2} \neq 0.
\label{log}
\end{equation}
Note that the non-transversality property is $\alpha$-independent.

This result can be checked by calculating explicitly the graviton self-energy.
In \cite{medrano}
one can find the result in the Feynmann gauge ($\alpha=-1$). We made the
calculation of
the logarithmic divergences for generic $\alpha$. If $\tilde\Pi_{\mu \nu,
\rho \sigma}^{\prime}$ denotes the
logarithmic divergences of the self-energy without propagators on external
legs,
the result is

\begin{eqnarray}
\tilde\Pi_{\mu \nu,
\rho \sigma}^{\prime}&=&\left\{\left({41\over 30}-{2\over 3}\beta
+\beta^2\right){\cal R}+
\left({13 \over 30}-{1\over 3}\beta+{1\over 2}\beta^2\right)k^2 {\cal Q}+
\left(-{27\over 80}+{1\over 4}\beta-{1\over 4}\beta^2\right)k^2 {\cal P}\right.
\nonumber\\&&
+\left. \left(-{59\over 240}+{1\over 12}\beta\right)k^4 {\cal T}+
 \left({27\over 40}-{1\over 2}\beta+{1\over 2}\beta^2\right)k^4 {\cal U}
\right\}{1\over 16 \pi^2} \hskip .1truecm {\rm ln} \hskip .05truecm
{\Lambda^2\over \mu^2},
\label{a41}
\end{eqnarray}

where $\beta=\alpha+1$, while
${\cal P}_{\mu \nu,\rho \sigma}= k_\mu k_\rho \eta_{\nu \sigma}+
k_\mu k_\sigma \eta_{\nu \rho}+k_\nu k_\rho \eta_{\mu \sigma}+
k_\nu k_\sigma \eta_{\mu \rho},
$ $
{\cal Q}_{\mu \nu,\rho \sigma}= k_\mu k_\nu \eta_{\rho \sigma}+
\eta_{\mu \nu} k_\rho k_\sigma,
$ $
{\cal R}_{\mu \nu,\rho \sigma}= k_\mu k_\nu k_\rho k_\sigma,
$ $
{\cal T}_{\mu \nu,\rho \sigma}= \eta_{\mu \nu} \eta_{\rho \sigma},
$ $
{\cal U}_{\mu \nu,\rho \sigma}= {1\over2}\left(\eta_{\mu \rho}
\eta_{\nu \sigma}+\eta_{\mu \sigma}\eta_{\nu \rho}\right).
$
It is easy to check that (\ref{a41}) verifies (\ref{log}).

Identity (\ref{ward}) is derived in the context of dimensional regularization.
Let us introduce the massive fields (the fields that are candidates to become
PV regulators). The dimensional cut-off permits to classify their
contributions to the logarithmic divergences. In presence of the
massive fields,
identity (\ref{ward}) looks formally the same. By subtracting
${2\over \alpha}k^\rho \tilde\Pi_{\mu \nu, \rho \sigma}$
in the two cases (with and without massive fields), we find that, setting
\begin{equation}
\tilde\Pi^{tot}_{\mu \nu, \rho \sigma}=\tilde\Pi_{\mu \nu, \rho \sigma}+
\tilde\Pi^{PV}_{\mu \nu, \rho \sigma},
\end{equation}
the PV contribution $\tilde\Pi^{PV}_{\mu \nu, \rho \sigma}$
is always transverse, i.e.\
\begin{equation}
k_\rho \tilde\Pi^{PV}_{\mu \nu, \rho \sigma}=0.
\label{pvtr}
\end{equation}
In conclusion, whichever regulator one uses (coupled to gravity in
a covariant way)
and whichever nonminimal coupling
one introduces, the logarithmic divergences brought by PV fields have not
the same structure as the logarithmic divergences that they should regularize.
We are lead to conclude that, even if the maximal divergences
($\delta^{(4)}(0)$)
can be regularized
independently of any gauge-fixing
(indeed they do not depend on the gauge-fixing \cite{anselmi,anselmi2}),
the nonmaximal divergences,
such as the logarithmic ones in dimension four,
cannot be regularized with each gauge-fixing.

One can wonder if this situation can be sensibly changed by replacing the
linear gauge-fixing ${G_F}_\mu=\partial^\nu \phi_{\mu \nu}$
with a nonlinear one, for example,
\begin{eqnarray}
{G_F}_\mu&=\partial^\nu \phi_{\mu \nu}+\phi_{\mu \nu}
(A\partial_\rho \phi^{\rho \nu}+
B \partial^\nu \phi)
+(C \eta^{\rho \nu} \phi+D \phi^{\rho \nu})\partial_\rho \phi_{\mu \nu}
\nonumber\\&
+\partial_\mu (E\phi^2+ F \phi_{\alpha \beta}\phi^{\alpha \beta}),
\end{eqnarray}
where $\phi=\phi^{\mu\nu}\eta_{\mu\nu}$.
It can be shown, by explicit computation,
that it is possible to fix the unknowns
{\it A, B, C, D, E} and $F$ in order to assure the transversality of the
logarithmic divergences of the graviton self-energy. However the solution is
not particularly enlightening and so we turn at once
to analyse what information can be derived from the properties of
the background field method. Indeed, the background field method gives a
simple way to avoid the obstacle that we have described in this section.

\section{Pauli-Villars one loop regularization}
\label{regula}

In this section we show that quantum gravity at the one loop order can be
regularized with one complex PV vector $W_\mu$ and one real PV tensor
$T_{\mu\nu}$.
The gauge-fixing is the unusual gauge-fixing that assures the identity
of the effective action with the effective action that one
finds with the method of the background field. This is not too
surprising since we want a {\sl covariant} regularization and
it is well known that the conventional gauge-fixings do not give
an off-shell covariant effective action. Moreover, a restriction on the form
of the gauge-fixing must be imposed, as a consequence of the reasonings
of the previous section.

The PV regulators are coupled in a nonminimal way. All the nonminimal
couplings listed in equations (\ref{nonmin}) and (\ref{nonmin2}) are
present. The complex vector can be
conventionally considered as the regulator of the ghosts, the real tensor
can be considered as
the regulator of the metric tensor. However, this is
no more than a conventional definition
in the sense that it is not true that the PV vector $W_\mu$
regularizes the divergences due to the ghosts and the PV tensor
$T_{\mu\nu}$ regularizes
the divergences due to the metric tensor
(this holds only when using the background field method).
It is anyway true that the two
regulators altogether regularize the total divergences.

The regularized lagrangian is
\begin{equation}
{\cal L}={\cal L}_g+{\cal L}_{gf}+{\cal L}_{gh}+\sum_{i=1}^{n_v}
{\cal L}_{v_i}+\sum_{i=1}^{n_t}{\cal L}_{t_i}.
\label{lalagra}
\end{equation}
${\cal L}_g$ is the Einstein lagrangian (in this section we set $\kappa=1$
for simplicity)
\begin{equation}
{\cal L}_g=-2\sqrt{-g}g^{\mu\nu}R_{\mu\nu}.
\end{equation}
(It differs by an overall sign with respect to the convention of the
previous section).
${\cal L}_{gf}$ is the gauge-fixing lagrangian
\begin{equation}
{\cal L}_{gf}=-\sqrt{-{\bar g}} {\bar g}^{\mu\nu}{\bar g}^{\rho\sigma}
{\bar g}^{\alpha\beta}
\left({\bar D}_\rho g_{\mu\sigma}-{1\over 2}{\bar D}_\mu g_{\rho\sigma}\right)
\left({\bar D}_\alpha g_{\nu\beta}-{1\over 2}{\bar D}_\nu g_{\alpha\beta}
\right).
\label{gauge-fixing}
\end{equation}
Here $\bar g_{\mu\nu}$ is an external metric tensor, which must be
set equal to the ``classical field'' $(g_{cl})_{\mu\nu}$
(i.e.\ the variables of the effective action \cite{marcus}).
$\bar D$ is the covariant derivative with
respect to $\bar g_{\mu\nu}$.

${\cal L}_{gh}$ is the ghost lagrangian
\begin{equation}
{\cal L}_{gh}=\sqrt{-{\bar g}}\bar C_\mu \bar g^{\mu\nu}\bar g^{\rho\sigma}
\delta_{BRS}\left(\bar D_\rho g_{\nu\sigma}-{1\over 2}\bar D_\nu g_{\rho\sigma}
\right),
\end{equation}
where $\delta_{BRS} \, g_{\mu\nu}=g_{\mu\rho}\bar D_\nu C^\rho+
g_{\nu\rho}\bar D_\mu C^\rho+(\bar D_\rho g_{\mu\nu})C^\rho$
(and, of course, $\delta_{BRS}\, \bar g_{\mu\nu}=0$).

${\cal L}_{v_i}$ are the lagrangians of the Pauli-Villars vectors ${W_i}_\mu$
\begin{equation}
{\cal L}_{v_i}=\sqrt{-g}\bar {W_i}_\mu (\Box+M_{v_i}^2) {W_i}^\mu-\sqrt{-g}
{R^\mu}_\nu \bar {W_i}_\mu {W_i}^\nu
\end{equation}
and finally ${\cal L}_{t_i}$ are the lagrangians of the PV tensors
${T_i}_{\mu\nu}$
\begin{eqnarray}
{\cal L}_{t_i}&=&{1\over 4}\sqrt{-g} (2 {T_i}^{\mu\nu}
(\Box+M_{t_i}^2){T_i}_{\mu\nu}
-T_i(\Box+M_{t_i}^2)T_i+4 {T_i}_{\mu\nu}{T_i}_{\rho\sigma}
R^{\mu\rho\nu\sigma}\nonumber\\
&&-4{T_i}_{\mu\nu}{{T_i}^\nu}_\rho R^{\mu\rho}+4 T_i {T_i}_{\mu\nu}R^{\mu\nu}
-2 {T_i}_{\mu\nu}{T_i}^{\mu\nu}R+{T_i}^2 R).
\end{eqnarray}
$\Box$ denotes the covariant D'Alembertian $g^{\mu\nu}D_\mu D_\nu$.
The PV vector $W_\mu$
and the PV tensor $T_{\mu\nu}$ are
not in irreducible
representations of the Poincar\`e group. So, the tensor corresponds to
a spin two particle plus a spin one particle plus two scalars,
while the
vector corresponds to a complex spin one particle plus
a complex scalar.
$M_{v_i}$ are the masses of the
PV complex vectors ${W_i}_\mu$, $M_{t_i}$ those of the PV real
tensors ${T_i}_{\mu\nu}$. Each regulator is
provided with a set of PV coefficients. We denote with $c_{v_j}$
the PV coefficients of the vector and with $c_{t_j}$ those of the tensor.
The tensor has bosonic statistics and the PV coefficient $c_t$ of Eq.\
(\ref{pvcond}) is equal to $1$, while the vector has fermionic statistic and
its PV coefficient $c_v$ is equal to $-2$.
Each set of PV coefficients $\{c_{v_j}\}$ and $\{c_{t_j}\}$
satisfy regularization conditions
of the kind (\ref{pvcond}) and the ``renormalization conditions''
that we shall discuss later on.

According to Ref.s \cite{anselmi,anselmi2}, the functional measure is
\begin{eqnarray}
d\mu &&\equiv
\prod_{\mu\leq\nu}{\cal D} [(-g)^{r-4\over 4r}g_{\mu\nu}]\,\,
\prod_\nu {\cal D} [(-\bar g)^{r-2\over 4r}\bar C_\nu]\,\,
\prod_\mu {\cal D} [(-g)^{r+2\over 4r}C^\mu]\nonumber\\
&&
\prod_{i=1}^{n_v}\left[
\prod_\nu {\cal D}^\prime [(-g)^{r-2\over 4r}\bar {W_i}_\nu]\,\,
\prod_\mu {\cal D}^\prime [(-g)^{r+2\over 4r}{W_i}^\mu]\right]\,
\prod_{j=1}^{n_t}\prod_{\mu\leq\nu}{\cal D}^\prime
[(-g)^{r-4\over 4r}{T_j}_{\mu\nu}]\nonumber\\&&
=\left({g\over \bar g}\right)^{r-2\over 4}
\prod_{\mu\leq\nu}{\cal D}g_{\mu\nu}\,
\prod_\nu {\cal D}\bar C_\nu\,
\prod_\mu {\cal D}C^\mu\,
\prod_{i=1}^{n_v}\left[
\prod_\nu {\cal D}^\prime\bar {W_i}_\nu\,
\prod_\mu {\cal D}^\prime {W_i}^\mu\right]\,
\prod_{j=1}^{n_t}\prod_{\mu\leq\nu}{\cal D}^\prime
{T_j}_{\mu\nu}.\nonumber\\
\label{mis}
\end{eqnarray}
The product over space-time points is understood.
Note that the antighost $\bar C_\mu$ has a power of $-\bar g$ and not
of $-g$, due to the fact that it is a vector
under background reparametrizations, but not under
BRS transformations, because the BRS transformation of the antighost
is the gauge-fixing \cite{anselmi,anselmi2}.

As shown in Ref.\ \cite{anselmi2}, the regularization can be
equivalently performed
by means of a unitary nonultralocal measure $d\mu_{PV}$.
It is obtained by integrating
away the PV regulators according to formula (\ref{dtnp}).
This is possible as far as the PV lagrangian is quadratic in the PV fields.
Unitarity is not affected, since the masses of the integrated fields are
arbitrarily large.
One has, after some manipulations,
\begin{eqnarray}
d\mu_{PV} &=&
\prod_{\mu\leq\nu}{\cal D} [(-g)^{r-4\over 4r}g_{\mu\nu}]\,\,
\prod_\nu {\cal D} [(-\bar g)^{r-2\over 4r}\bar C_\nu]\,\,
\prod_\mu {\cal D} [(-g)^{r+2\over 4r}C^\mu]\cdot\nonumber\\
&&
{\rm det}_v\left[{{\cal O}^\mu}_\nu\right]^{-1}\,\,
{\rm det}_t\left[{{\cal O}^{\mu\nu}}_{\rho\sigma}\right]^{1\over 2},
\end{eqnarray}
where
\begin{equation}
{{\cal O}^\mu}_\nu=\delta^\mu_\nu \Box_v-{R^\mu}_\nu
\end{equation}
and
\begin{eqnarray}
{{\cal O}^{\mu\nu}}_{\rho\sigma}&=&{1\over 2}
(\delta^\mu_\rho\delta^\nu_\sigma+\delta^\mu_\sigma\delta^\nu_\rho)
(\Box_t-R)+{1\over 2}({{{R^\mu}_\rho}^\nu}_\sigma+{{{R^\nu}_\rho}^\mu}_\sigma
+{{{R^\mu}_\sigma}^\nu}_\rho+{{{R^\nu}_\sigma}^\mu}_\rho)
\nonumber\\&&
+g^{\mu\nu}R_{\rho\sigma}
+g_{\rho\sigma}R^{\mu\nu}
+{1\over 2-r}g^{\mu\nu}
(rR_{\rho\sigma}+g_{\rho\sigma}R)\nonumber\\&&
-{1\over 2}(R^\mu_\rho\delta^\nu_\sigma+R^\nu_\rho\delta^\mu_\sigma+
R^\mu_\sigma\delta^\nu_\rho+R^\nu_\sigma\delta^\mu_\rho).
\end{eqnarray}
$\Box_v$ is the covariant D'Alembertian acting on vectors,
$\Box_t$ is the one acting on symmetric tensors with two indices.
$\det_v$ and $\det_t$ are ``Pauli-Villars determinants'' \cite{anselmi2}.
Their formal definition is the following
\begin{equation}
{\rm det_{\it u}} [{\cal O}]^{{1\over 2}\sum_i c_{u_i}}\equiv
\lim_{\{M_{u_i}^2\}\rightarrow\infty}
\prod_{i=1}^{n_u} \hskip .1truecm
{\rm det}\left[{\cal O}+M_{u_i}^2\, 1_u\right]^{{1\over 2}c_{u_i}},
\label{PVdet}
\end{equation}
where $u=s$ for scalars, $v$ for vectors and $t$ for tensors,
while $1_u$ is $1$ for scalars, $\delta^\mu_\nu$ for vectors and
${1\over 2}(\delta^\mu_\rho\delta^\nu_\sigma+
\delta^\nu_\rho\delta^\mu_\sigma)$ for symmetric tensors. $n_u$ is the number
of copies of PV fields of type $u$.

We have chosen the gauge-fixing of formula (\ref{gauge-fixing})
both because the usual one cannot work, as explained in section \ref{9},
and because this is the gauge-fixing inherited from the approach
of the background field method. Indeed, in the context of the background
field method, it is simple to see that the divergences are
regularized.

Let us make a brief digression on a scalar in external gravity.
Consider the lagrangian
\begin{equation}
{\cal L}_s={1\over 2}\sqrt{-g}g^{\mu\nu}\partial_\mu \varphi
\partial_\nu\varphi.
\label{varphi}
\end{equation}
Surely, PV scalars $\chi_j$ with lagrangians
\begin{equation}
{{\cal L}_j}_{PV}={1\over 2}\sqrt{-g}g^{\mu\nu}\partial_\mu \chi_j
\partial_\nu\chi_j+{1\over 2}M_j^2\chi_j^2
\end{equation}
and PV coefficient $c=1$ (see (\ref{pvcond}))
regularize ${\cal L}_s$: expanding $g_{\mu\nu}$ with respect to flat space,
it is evident that
the propagator of $\chi_j$ is that of a
massive scalar and the couplings with gravity are the same as those of
$\varphi$. However, ${\cal L}_{PV}$ is not covariant. As a matter of fact,
the sum of the covariant lagrangians
\begin{equation}
{{\cal L}^\prime_j}_{PV}={1\over 2}\sqrt{-g}(g^{\mu\nu}\partial_\mu
\chi_j \partial_\nu\chi_j+M_j^2\chi_j^2)
\label{chi}
\end{equation}
also regularizes ${\cal L}_s$. Indeed, in the variables $\tilde \varphi=
(-g)^{1\over 4}\varphi$ and $\tilde \chi_j=(-g)^{1\over 4}\chi_j$
the lagrangians ${\cal L}_s$ and ${{\cal L}^\prime_j}_{PV}$ become
\begin{eqnarray}
\tilde{\cal L}_s&=&{1\over 2}\sqrt{-g}g^{\mu\nu}\partial_\mu
[(-g)^{-{1\over 4}}\tilde\varphi] \partial_\nu[(-g)^{-{1\over
4}}\tilde\varphi],
\nonumber\\
{\tilde{\cal L}}^\prime_{j\, PV}&=&{1\over 2}\sqrt{-g}g^{\mu\nu}\partial_\mu
[(-g)^{-{1\over 4}}\tilde\chi_j] \partial_\nu [(-g)^{-{1\over 4}}\tilde\chi_j]+
{1\over 2}M_j^2\tilde\chi_j^2.
\end{eqnarray}
In these variables it is clear that
$\sum_j{\tilde{\cal L}}^\prime_{j\, PV}$ regularizes
$\tilde{\cal L}_s$.
Now we can go back to ${\cal L}_s$ and ${{\cal L}_j^\prime}_{PV}$
with the inverse change of variables and conclude
that $\sum_j{{\cal L}_j}_{PV}^\prime$
regularizes ${\cal L}_s$. The change of variables can only
affect the $\delta(0)$
divergences, but we know that they are in any case correctly arranged
by the functional measure \cite{anselmi}.

Let us now see what happens in the case of gravity.
We prove that the lagrangian (\ref{lalagra}) regularizes
the one loop divergences.
The background field method has the feature that, at the one loop
order, the problem is reduced to a problem of fields in {\sl external}
gravity (the background) and so the above reasoning can be applied.
Let $g_{\mu\nu}=\bar g_{\mu\nu}+h_{\mu\nu}$ be the decomposition of the metric
tensor $g_{\mu\nu}$ into the background part $\bar g_{\mu\nu}$
plus the quantum fluctuation $h_{\mu\nu}$.
The sum of the Einstein lagrangian ${\cal L}_g$
plus the gauge-fixing term ${\cal L}_{gf}$ reduces to
\cite{sagnotti}
\begin{eqnarray}
{\cal L}_{(bf)}&=&{1\over 4}\sqrt{-\bar g} (2 {h}^{\mu\nu}\Box {h}_{\mu\nu}
-h\Box h+4 h_{\mu\nu}h_{\rho\sigma}
\bar R^{\mu\rho\nu\sigma}\nonumber\\
&&-4h_{\mu\nu}{h^\nu}_\rho \bar R^{\mu\rho}+4 h\, h_{\mu\nu}\bar R^{\mu\nu}
-2 h_{\mu\nu}h^{\mu\nu}\bar R+h^2 \bar R),
\end{eqnarray}
where the indices are raised and lowered by means of the background
metric tensor.
Now $\Box$ denotes the D'Alembertian in the background.
The ghost lagrangian ${\cal L}_{gh}$ reduces to
\begin{equation}
{\cal L}_{gh(bf)}=\sqrt{-\bar g}\bar C_\mu \Box C^\mu-\sqrt{-\bar g}
{\bar R^\mu}_{\phantom{.}\nu} \bar C_\mu C^\nu.
\end{equation}
In the PV lagrangians ${\cal L}_{v_i}$ and ${\cal L}_{t_i}$, one can simply
substitute $g_{\mu\nu}$ with $\bar g_{\mu\nu}$.
Let us denote by ${\cal L}_{v_i(bf)}$ and
${\cal L}_{t_i(bf)}$ the PV lagrangians obtained in this way.
This gives all that matters for the application of the background field
method at one loop.
The similarities between
${\cal L}_{(bf)}$ and ${\cal L}_{t_i(bf)}$ and between ${\cal L}_{gh(bf)}$
and ${\cal L}_{v_i(bf)}$ are evident.
Indeed, $\sum_i{\cal L}_{t_i(bf)}$ regularizes ${\cal L}_{(bf)}$, while
$\sum_i{\cal L}_{v_i(bf)}$ regularizes ${\cal L}_{gh(bf)}$.
To make this clearer,
let us introduce the background vielbein $\bar e_\mu^a$
and make the change of variables $\tilde h_{ab}=h_{\mu\nu}\bar
e^\mu_a \bar e^\nu_b$ and $\tilde {\bar C}_a=\bar C_\mu \bar e^\mu_a$,
$\tilde C^a= C^\mu \bar e^a_\mu$. Then ${\cal L}_{(bf)}$
and ${\cal L}_{gh(bf)}$
become, respectively,
\begin{eqnarray}
\tilde{\cal L}_{(bf)}&=&{1\over 4}\sqrt{-\bar g} (2 {\tilde h}^{ab}\Box {
\tilde h}_{ab}
-\tilde h\Box \tilde h+4 \tilde h_{ab}\tilde h_{cd}
\bar R^{acbd}\nonumber\\
&&-4\tilde h_{ab}{\tilde h^b}_c \bar R^{ac}+4 \tilde h\, \tilde h_{ab}
\bar R^{ab}
-2 \tilde h_{ab}\tilde h^{ab}\bar R+\tilde h^2 \bar R),\nonumber\\
\tilde{\cal L}_{gh(bf)}&=&
\sqrt{-\bar g}\tilde {\bar C}_a \Box \tilde C^a-\sqrt{-\bar g}
{\bar R^a}_b \tilde {\bar C}_a \tilde C^b.
\end{eqnarray}

A similar change of variables can be done for $\bar {W_i}_\mu$,
${W_i}_\mu$ and ${T_i}_{\mu\nu}$. At the end we reduce to a set of cases
that are analogous to the scalar $\varphi$ and the regulator $\chi$.
This proves that the one loop regularization works.

It is now clear why, when using the ordinary method (non background), the $W$'s
cannot be considered as the regulators of the ghosts and the $T$'s
as the regulators of the metric tensor $g_{\mu\nu}$. Indeed, only in
formalism of
the background field method one can say that the lagrangian of $h_{\mu\nu}$
corresponds to the lagrangians of ${T_i}_{\mu\nu}$
and the lagrangian of $\bar C_\mu$, $C^\mu$ to those of
$\bar {W_i}_\mu$, ${W_i}^\mu$, in the same way as the lagrangian of
$\chi$ (\ref{chi}) corresponds to that of $\varphi$ (\ref{varphi}).
Moreover, ${\cal L}_{v_i}$ is covariant with respect to $g_{\mu\nu}$,
while ${\cal L}_{gh}$ is not. It is nevertheless true, of course,
that the regulators $W$ and $T$ altogether regularize the total divergences.

\section{Pauli-Villars one loop renormalization}
\label{renorma}

The further conditions that
one has to impose on the PV coefficients $c_{u_j}$ of each regulator-type $u$
in order to renormalize the one loop divergences
without explicit introduction of counterterms are
\begin{equation}
\sum_j c_{u_j} (M_{u_j}^2)^p \left({\rm ln} \hskip .05truecm
{M_{u_j}^2\over\mu_u^2}\right)^
{\epsilon_r}=0,
\label{logcond2}
\end{equation}
where $\epsilon_r=1$ and $p$ takes the values
$0,1,2\ldots {r\over 2}$ if $r$ is even, while
$\epsilon_r=0$ and $p$ takes the values
${1\over 2},{3\over 2},\ldots {r\over 2}$ if $r$ is odd.
Eq.\ (\ref{logcond2}) contains also Eq.\ (\ref{logcond}).
As a matter of fact, using the background field method,
the Pauli-Villars regularization corrects
any one loop diagram by adding to it a sum of similar diagrams in which
the PV fields circulate. It is not difficult to
get convinced that the divergent expressions in the PV masses can only
have the form of the left hand side of Eq.\ (\ref{logcond2}).
In the appendix we exhibit the standard calculation that one encounters.
Notice that, using the ordinary formalism (non-background) only the sum of the
diagrams of a given order is well regularized and not the single diagrams.
The total set of conditions that must be satisfied by the PV coefficients
are thus Eq.s (\ref{pvcond}) and (\ref{logcond2}).
They permit to make the one loop divergences disappear without
any explicit introduction of counterterms.

Now and in the following section we discuss Eqs. (\ref{logcond2}) under various
and different respects. We shall see what happens when all of them are imposed,
as well as when only some of them are imposed.
We shall omit the index $u$
that specifies the kind of regulator, for now.

There is only one equation which depends on
the scale $\mu$, namely the equation with $p=0$ when $r$ is even.
For any other $p$ the scale $\mu$ is immaterial due to
(\ref{pvcond}). For $p=0$, varying $\mu$ is equivalent
to substitute the condition
\begin{equation}
\sum_j c_j \hskip .1truecm {\rm ln} \hskip .05truecm {M_j^2\over\mu^2}=0
\label{pauli}
\end{equation}
with the condition
\begin{equation}
\sum_j c_j \hskip .1truecm{\rm ln} \hskip .05truecm {M_j^2\over\mu^2}
={\rm const.}
\label{paulipop}
\end{equation}
This fact is immediate consequence of the first equation of
(\ref{pvcond}).
Clearly, there is no reason for preferring one scale to any other.
A similar
arbitrariness is usually related to the subtraction
of a divergent term:
a counterterm renormalizes a coupling constant and one needs to
specify a normalization condition, to
fix the value of the coupling constant at some energy.
However, within our formalism, we never  introduce counterterms,
nevertheless we still get an arbitrariness. The counterterms that
are avoided thanks to (\ref{pauli}) are quadratic
in the Riemann tensor.
In the ordinary approach, one is forced to introduce them
by the need of getting a
convergent one loop effective action\footnotemark\footnotetext{
We recall that the finiteness of quantum gravity at the one loop order
in dimension four and in absence of matter, is immaterial to our reasonings,
which can be applied in any dimension (greater than two) and also when matter
is present. We shall return later back to this point.}.
The introduction of these counterterms
is equivalent to the introduction of new coupling constants,
i.e.\ to a substantial modification of the starting lagrangian.
Then, the arbitrariness is fixed
by normalizing these new coupling constants (and not the
Newton constant). Within our formalism, instead,
we remain with {\sl two} arbitrary constants
(the scales $\mu_v,\, \mu_t$ of the
vector $W$ and the tensor $T$), and
there are no couplings which can make them
disappear after a suitable normalization.
In the next section we shall explicitly discuss the possibility that
$\mu_u$ are to be fixed experimentally.

Setting $M_j^2=t_j \lambda^2$, where $\lambda$
is a massive parameter to be
sent to infinity and $t_j$ are constants to be suitably chosen,
the system made of Eq.s (\ref{pvcond})
and (\ref{logcond2}) admits a solution in the unknowns $c_j$
if and only if (from now on we shall only write the formul\ae\ for $r$ even)
\begin{equation}
{\rm det} \left( \begin{array}{cccc}
1 & \ldots & \ldots & 1 \\
{\rm ln} \hskip .05truecm t_1 & \ldots & \ldots &
{\rm ln}\hskip .05truecm t_n \\
t_1 & \ldots & \ldots & t_n \\
t_1\hskip .1truecm{\rm ln}\hskip .05truecm t_1 & \ldots & \ldots & t_n
\hskip .1truecm
{\rm ln}\hskip .05truecm t_n \\
t_1^2 & \ldots & \ldots & t_n^2 \\
t_1^2\hskip .1truecm{\rm ln}\hskip .05truecm t_1 & \ldots & \ldots &
t_n^2\hskip .1truecm{\rm ln}\hskip .05truecm t_n \\
\ldots & \ldots & \ldots & \ldots \\
t_1^{r/2} & \ldots & \ldots & t_n^{r/2} \\
t_1^{r/2} \hskip .1truecm  {\rm ln} \hskip .05truecm  t_1
& \ldots & \ldots & t_n^{r/2}
\hskip .1truecm  {\rm ln} \hskip .05truecm  t_n
\end{array}\right) \neq 0.
\label{dim4}
\end{equation}
This will be generally true if one chooses $t_i \neq t_j$
for $i \neq j$.
We see that the number $n$ of PV fields that are required for making our
procedure work is $2r$, i.e.\ two times the number of PV fields that are
usually needed.
Indeed, in the ordinary Pauli-Villars regularization one only imposes
(\ref{pvcond}).

The next remark regards the dependence of the PV coefficients $c_j$
on $\lambda$. Eq.\ (\ref{logcond2}) gives
\begin{equation}
\sum_j c_j \hskip .1truecm t_j^{p} \hskip .1truecm {\rm ln}
\hskip .05truecm t_j=0
\label{dime}
\end{equation}
for $p\neq 0$ and
\begin{equation}
\sum_j c_j \hskip .1truecm {\rm ln} \hskip .05truecm t_j=
- c\hskip .1truecm {\rm ln} \hskip .05truecm
{\lambda^2\over \mu^2}
\label{pauli2}
\end{equation}
for $p=0$. One can decide to impose (\ref{dime}) only (together
with (\ref{pvcond})) and not (\ref{pauli2}). In that case, the PV coefficients
$c_j$ do not depend on ${\lambda\over \mu}$.
This situation has some similarity with what happens
in dimensional regularization, in the following sense.
Due to the properties of the dimensional
regularization technique, only the logarithmic divergences
are nontrivial (the poles).
Correspondingly, Eq.\ (\ref{dime}) makes all the one loop
divergences vanish except for the logarithmic ones.
Then, the logarithmic divergences
in the PV masses are those proportional
to the left hand side of Eq.\ (\ref{pauli}).
The fact that the non-logarithmic
divergences can be made disappear by simply fixing
some numerical coefficients (the coefficients $c_j$) means that
they are often non-meaningful divergences, at least
in a renormalizable theory, and
dimensional regularization shows it with evidence. In quantum gravity,
however, the non-logarithmic divergences
can be important, not only for a correct definition of
the functional integral,
as we shall discuss in the next section. Moreover, even if the dimensional
regularization
is a good computational technique at the level of Feynmann diagrams,
it may be useful to possess a good regularization that is
definable at the level of the functional integral.

In the context of our regularization scheme, one {\sl can} impose
Eq.\ (\ref{pauli}) (or equivalently (\ref{pauli2})).
In this case, the logarithmic divergences also disappear,
but the coefficients $c_j$ are no longer independent of
${\lambda^2\over \mu^2}$.
Indeed, they
behave as ${\rm ln} \hskip .05truecm {\lambda^2\over \mu^2}$
when ${\lambda^2\over \mu^2}$ goes to infinity.
This means that the imposition of Eq.\ (\ref{pauli}) corresponds to
a kind of renormalization. The important fact that we want to stress is that
we do not need to introduce counterterms in order to get this result.
The divergences are renormalized by the PV fields themselves. In a
renormalizable theory there is no problem in introducing
the explicit counterterms
that cancel the logarithmic divergences, because they are
of the same form as the terms of the starting lagrangian,
as a consequence of the adimensionality of the coupling constant.
So, the fact that they do not
appear explicitly in the context of a suitable PV regularization, like ours,
is not upsetting. However, in the case of gravity the counterterms
that should cancel the logarithmic divergences
are not of the same form as the terms of the starting lagrangian,
because they are quadratic in the Riemann tensor.
So, a procedure that avoids their explicit introduction
is deserving of interest. As a matter of fact, this was our main motivation
for searching for a formulation of the covariant PV regularization
of quantum gravity at the one loop order.

The logarithmic behaviour of the PV coefficients is the only one that is
permitted if they have to diverge and the theory has to remain meaningful.
Indeed, when computing Feynmann diagrams,
one is interested in the limit when the PV masses tend to infinity.
So, one expands in powers of the masses and finds
expressions of the form
\begin{equation}
\sum_j c_j \left({1\over M_j^2}\right)^p
\end{equation}
or, at worst, of the form
\begin{equation}
\sum_j c_j \left({1\over M_j^2}\right)^p \hskip .1truecm {\rm ln}
\hskip .05truecm {M_j^2\over \mu^2}
\end{equation}
for $p\geq 1$.
These expressions tend to zero even if the $c_j$ are logarithmic in
$\lambda^2\over \mu^2$.
There seem to be no need for the $c_j$ either to be
constants or to tend to constants.
Instead, if the $c_j$ tended to infinity as powers of $\lambda$,
there would be further divergences to be renormalized\footnotemark
\footnotetext{In the article of
W.\ Pauli and F.\ Villars \cite{paulivillars}
some similar remarks about the behaviour of $c_j$ can be found.
}.

Let us now make some heuristic remarks that may facilitate the
comparison between our
regularization technique and the familiar Pauli-Villars technique.

The ordinary PV regularization is equivalent to the
substitution of the propagator, say ${1\over k^2}$ for simplicity,
with
\begin{equation}
{1\over k^2}-\sum_j c_j{1\over k^2 - M_j^2}.
\label{propa}
\end{equation}
We see that, being $c_j\sim {\rm ln} \;{\lambda^2\over \mu^2}$ and $M_j\sim
\lambda^2$,
the modified propagator reduces to the usual one as
${\lambda^2\over \mu^2}\rightarrow\infty$, although the $c_j$ diverge.
At the same time, this argument shows that one could hardly
accept eventual conditions on the $c_j$ that make them tend to
infinity as powers of $\lambda^2\over \mu^2$.

Let us see how similar formal arguments are applied to the functional
integral. In the usual formalism for PV fields (non DTNP) the action
of the PV fields is of the kind (see, for example, \cite{diagrammar})
\begin{equation}
{1\over c_j}\chi_j (\Box + M_j^2)\chi_j+ {\cal L}_{int},
\end{equation}
where $\Box$ is the flat D'Alembertian and
${\cal L}_{int}$ denotes a generic interaction. If one formally lets
$\lambda^2\over \mu^2$ tend to infinity, the action behaves as
\begin{equation}
{\lambda^2\over {\rm ln}\;{\lambda^2\over \mu^2}}\chi_j^2.
\end{equation}
Thus, a large ${\lambda^2\over \mu^2}$ imposes $\chi_j\equiv 0$, as usual.

In the DTNP
formalism, on the other hand, the action is
\begin{equation}
\chi_j (\Box + M_j^2)\chi_j+ V\chi_j^2,
\end{equation}
where $V\chi^2_j$ is a generic vertex,
but now it is the definition of functional integral that changes.
If we integrate in $\chi_j$ we formally obtain
\begin{equation}
\det (\Box + M_j^2+V)^{c_j\over 2}=\det (\Box+M_j^2)^{c_j\over 2}
\cdot \det \left(1+{V\over \Box +M_j^2}\right)^{c_j\over 2}
\label{parapa}
\end{equation}
The first factor arranges the correct normalization of the functional
integral \cite{anselmi2}. The second factor tends formally to one.

However, we
must note that there is an important difference between the
usual substitution of the propagator with one of the form (\ref{propa})
(this method regularizes at any loop) and the one loop
technique (sometimes known as Pauli-Bethe). The DTNP approach is
a variant of the latter technique. The one loop
technique regularizes
one loop diagrams by adding other diagrams in which
the massive regulators circulate. On the other hand,
when replacing the propagator with one of the form
(\ref{propa}), the most simple one loop divergences are proportional to
expressions of the form
\begin{equation}
\sum_{ij}c_i\, c_j \, A(M_i,M_j).
\label{cc}
\end{equation}
In general, one has a product of as many coefficients $c_j$
as propagators making the loop. Expression (\ref{cc}) corresponds
to a self-energy, i.e.\ two propagators.
In general, $A(M_i,M_j)$ does not depend trivially on the masses.
Consequently, to require the above expression (or similar
expressions) to be zero (in order to make
the PV fields eat their own divergences) would mean
to impose a quadratic equation on the PV coefficients $c_j$,
thus lacking the simplicity of conditions (\ref{logcond2})
and the consequent analysis on the behaviour of the $c_j$.
Thus, the one loop approach is very preferable to the most
general approach.
It does not seem simple to find a generalization of the Pauli-Bethe
technique or the DTNP approach beyond one loop, nevertheless
it is worth thinking about it.

At this point, it may be helpful to
summarize the properties of the covariant Pauli-Villars
regularization of quantum gravity at the one loop order.

First of all, it is possible to make the maximal divergences disappear
\cite{anselmi,anselmi2}. If one uses a naive cut-off,
one verifies that they do not vanish,
not even with Fujikawa's measure, save the theory is
supersymmetric or, in any case, contains a null total number of
degrees of freedom (with the convention that
the fermionic degrees of freedom contribute with a negative sign).
On the other hand, in dimensional
regularization, the maximal divergences are trivially absent.
Turning to
a Pauli-Villars regularization, one can
make the $\delta(0)$ divergences vanish
and contemporarily eliminate also the maximal divergences in the PV masses,
by suitably fixing the PV coefficients $c_j$ (condition (\ref{logcond})).

Secondly, the nonmaximal divergences, in
particular the logarithmic ones, require the introduction of counterterms
different from the terms of the original lagrangian,
even in the context of dimensional regularization.
The Pauli-Villars approach,
instead, permits to make the PV fields produce the desired counterterms
by themselves, with no modification of the original lagrangian.
Such a result can be achieved by adding conditions on the coefficients $c_j$
(see (\ref{logcond2})).
This property is immaterial
in ordinary renormalizable quantum field theories,
but could be very interesting in quantum gravity.

One can wonder what about an ordinary nonrenormalizable theory. It seems that
it is possible to make it finite at one loop in any case,
without introducing counterterms of a form different from the form of
the terms of the starting lagrangian. This is true but the price,
as already noted, is the introduction
of arbitrary constants $\mu$.
Were it possible to generalize the procedure at any loop,
it would seem possible to make any theory finite at any desired
number of loops. However, this would not save the increasing arbitrariness
that would appear loop by loop. To eliminate the divergences that appear
at the $k^{th}$ loop order, one would presumably
need something like $r(k+1)$ PV fields to satisfy all the conditions
on the PV coefficients and
$2k$ arbitrary scales $\mu$ would appear as a consequence.
This should be the way in which, in our procedure,
a nonrenormalizable
theory exhibits the infinite arbitrariness that makes it unphysical.

However, these remarks about nonrenormalizable theories are valid in general.
One can still hope that a {\sl particular} nonrenormalizable theory can
be made finite with a finite number of conditions of the kind (\ref{logcond2})
and consequently with a finite number of PV fields.
For example, this could still be the case of quantum gravity.
Thus the correct way of defining quantum gravity
could be a combination of renormalization and finiteness, i.e.\
it could be possible to make the theory finite only after having
made the divergences vanish up to some loop order in the way that we have
seen. Then one should be able to prove that all the divergences that survive
at the subsequent orders are zero on shell
(these divergences should be regularized with an ordinary technique,
but one should take into account the presence and the contributions
of the PV fields).
In this way only a finite arbitrariness would survive.

It is straightforward to note that with the same
technique that we have exhibited, it is possible to make the
one loop divergences to vanish when gravity is coupled to matter.
When matter is present, the usual treatment of quantum gravity
shows that it is not even finite at the one loop order
in dimension four.
Let the lagrangian of a scalar $\varphi$ of mass $m$ be
\begin{equation}
{\cal L}_s={1\over 2}\sqrt{-g}(g^{\mu\nu}\partial_\mu \varphi\partial_\nu
\varphi+m^2\varphi^2).
\label{lagra1}
\end{equation}
The Pauli-Villars regularization at the one loop order is performed by PV
scalars $\chi_j$, with square masses $M^2_{t_j}+m^2$ and PV coefficients
$c_{t_j}$: the square mass is that of the PV regulator ${T_j}_{\mu\nu}$
plus the square mass of $\varphi$, while the
PV coefficient is exactly the same as that of ${T_j}_{\mu\nu}$.
The regularized lagrangian is
\begin{eqnarray}
{\cal L}_{reg}&=&{1\over 2}\sqrt{-g}(g^{\mu\nu}\partial_\mu \varphi\partial_\nu
\varphi+m^2\varphi^2)+
{1\over 2}\sum_j^{n_t}\sqrt{-g}(g^{\mu\nu}\partial_\mu \chi_j\partial_\nu
\chi_j+(M_{t_j}^2+m^2)\chi_j^2)\nonumber\\&&
+{1\over 2}\sum_j^{n_t}\sqrt{-g}(g^{\mu\nu}T_j
-2T_j^{\mu\nu})\partial_\mu\varphi\partial_\nu\chi_j\nonumber\\&&
-{1\over 16}\sum_j^{n_t}\sqrt{-g}(g^{\mu\nu}(2T^{\rho\sigma}_j
{T_j}_{\rho\sigma}-T_j^2)+4 T^{\mu\nu}_jT_j-8 {T_j^\mu}_\rho T_j^{\rho\nu})
\partial_\mu\varphi\partial_\nu\varphi\nonumber\\&&
+{1\over 2} m^2\sum_j^{n_t}\sqrt{-g}
\left(T_j\varphi\chi_j+{1\over 8}\varphi^2(T_j^2-2 T_j^{\mu\nu}{T_j}_{\mu\nu})
\right).
\label{lagra2}
\end{eqnarray}
${\cal L}_{reg}$ is determined by demanding any one loop diagram
corresponding to ${\cal L}_s$ to be corrected by a sum of similar
diagrams with loops of PV regulators. ${\cal L}_{reg}$ can be easily found
by thinking to what happens in the context
of the background field method. One starts from lagrangian
(\ref{lagra1}) and re-writes it according to the
rules of the background field method, keeping only those terms
that are interesting for the one loop diagrams. Then one
regularizes this background lagrangian, with suitable PV regulators,
thus finding a regularized background lagrangian.
Finally, one looks for a regularized {\sl non}-background lagrangian.
It is determined by requiring that it reduces to
the regularized background one when using the
method of the background field. This gives (\ref{lagra2}), that should be
added to (\ref{lalagra}), and
implies that the PV coefficients of the
$\chi$-regulators are equal to those of the $T$-regulators, while the square
mass of $\chi_j$ is the sum of the square mass of $T_j$ and the square
mass of $\varphi$.

Consider the diagrams with a
loop made both by scalars $\varphi$ and gravitons $h_{\mu\nu}$. If
$\chi$ and $T_{\mu\nu}$ satisfied independent regularization conditions,
this kind of diagrams would not be regularized, in general.
That is why the masses and PV coefficients of the $\chi$-regulator are not
independent of those of the $T$-regulator.
As a matter of fact,
the $\chi$- and $T$-regulators should be considered as a whole, as well
as the fields $h_{\mu\nu}$ and $\varphi$.
In this way, no new scale $\mu$ is introduced
when matter is coupled to gravity.

Instead, the fact that the ghosts have their own scale $\mu_v$
(i.e.\ the scale associated to $W$, which is not related to
the scale $\mu_t$ associated to $T$), is due to the fact that
the physical diagrams cannot contain loops
made partly by ghosts and partly by gravitons.
These diagrams would have external ghost legs and we do not require to
regularize them. Indeed,
demanding the regularization of these kind of diagrams (it could
be convenient for applications of Ward identities) would
produce, following the above scheme of reasoning, a non-BRS-invariant
regularized lagrangian. This is something that we want to avoid.

We conclude this section with a comment about treating
a renormalizable quantum field theory with our technique.
As a matter of fact,
one can easily convince oneself that on such a theory our procedure
has the same effect, at one loop, as any ordinary one.
Consider, for example, the vertex amplitude $\Gamma^{(4)}(p)$
for the theory $\lambda \phi^4$ in four dimensions
at the one loop order
($p$ denoting collectively the momenta on which
 $\Gamma^{(4)}$ depends).
We want to compare the results that one obtains with our procedure,
represented by conditions (\ref{logcond}) and
(\ref{logcond2}), with the results
that one obtains with an ordinary procedure, which for convenience
we choose to be represented by the same formal set-up, where only
conditions (\ref{logcond}) are imposed,
while (\ref{logcond2}) are not imposed. In other
words, we want to check that the extra conditions
(\ref{logcond2}) have no physical
effect on a renormalizable field theory.
The ordinary procedure leads to a result
of the form
\begin{equation}
\Gamma^{(4)}(p)=i\lambda\left[Z_1+\lambda\left({\alpha\over \varepsilon}
+\Gamma(p)\right)\right],
\end{equation}
where $1\over \varepsilon$ stands for the expression
for $\sum_jc_j\, {\rm ln}\, {M_j^2\over \mu^2}$, while
$Z_1$ is the vertex
renormalization constant, $\alpha$ is a number and $\Gamma(p)$ is a
suitable finite function of the momenta. One can fix $Z_1$ by choosing
a normalization condition like, for example,
$\Gamma^{(4)}(\bar p)=i\lambda$,
$\bar p$ being some reference scale. At the end one gets
\begin{equation}
\Gamma^{(4)}(p)=i\lambda[1+\lambda(
\Gamma(p)-\Gamma(\bar p))].
\label{bell}
\end{equation}
Our scheme, on the other hand, is equivalent
to replace $1\over \varepsilon$
with some constant [compare with (\ref{pauli}) and (\ref{paulipop})],
that here we call $\beta$, so that we have
\begin{equation}
\Gamma^{(4)}(p)=i\lambda[Z_1+\lambda(
\beta+\Gamma(p))].
\end{equation}
As before, we set
$\Gamma^{(4)}(\bar p)=i\lambda$, thus recovering expression (\ref{bell}),
exactly as in the ordinary approach. In the case of gravity, the difference
is that there is no $Z_1$ that can absorb $1\over \varepsilon$ or $\beta$,
since the divergent terms (quadratic in the Riemann tensor) correspond
to no coupling constant $\lambda$ in the original action.

What about higher loop effects of our technique on a
renormalizable theory?
In absence of a generalization of our method beyond one loop, the best
thing that we can do is to
combine it with an ordinary method (the dimensional technique, for instance),
remembering that we have to take into
account the contribution of the PV fields.
The problem is then whether this contribution is trivial or not.
In fact, we can show that in such a double regularized scheme,
the massive PV regulators do not contribute at all, by the
following simple argument. The ordinary regulator that one introduces to give
meaning to the theory beyond one loop is sufficient by itself to make the
theory completely meaningful. So, there is no problem in taking the limit
of infinite PV masses, since
the theory is well-behaved. Suppose we take this limit directly on the
functional integral. The reasonings that we made around formul\ae\
(\ref{propa})-(\ref{parapa}) are no more formal and heuristic:
in the case we are dealing
with, they are rigorous and they show that one can put the PV regulators
identically equal to zero
(even if the $c_j$ diverge logarithmically with the PV masses, as we noticed).
Thus there is no difference between this doubly regularized technique
and the usual ones when the theory is
renormalizable. On the other hand, the analogous argument cannot
be repeated for a theory like Einstein gravity, since there exists no
ordinary regulator that is able to give meaning to it: indeed, any ordinary
regulator, due to the need of introducing counterterms that are quadratic
in the Riemann tensor, inevitably turns Einstein
gravity into higher derivative
gravity. We recall that in this paper our attention is concentrated on
Einstein gravity and on no other kind of gravity.

\section{Renormalization of wave functions and coupling constants}
\label{coupling}

Let us now analyze the possibilities offered by our one loop regularization
from
a slightly different point of view. Suppose we only impose the regularization
conditions (\ref{pvcond}) for now. We want to discuss
what renormalization conditions we
are {\sl forced} to introduce, avoiding the imposition of conditions
that are not strictly required. This is the attitude of the present section.
We shall
specialize to the case of dimension four, for simplicity. We first discuss
pure gravity and then, to point out that the fact that the finiteness of
pure gravity in dimension four is immaterial to our purposes, we shall repeat
the reasonings for gravity coupled to matter (a massless scalar field).

Suppose the PV masses are taken to be finite, for now, and
consider the terms of the one loop effective action $\Gamma_{eff}$
that are expected to diverge when one lets the PV masses
go to infinity.
In dimension four these terms can only be proportional to
$\Lambda^4_u\equiv\sum_jc_{u_j}\, M_{u_j}^4\, {\rm ln} (M_{u_j}^2/\mu_u^2)$,
$\Lambda^2_u\equiv\sum_jc_{u_j}\, M_{u_j}^2\, {\rm ln} (M_{u_j}^2/\mu_u^2)$ and
${\rm ln}\Lambda_u^2/\mu_u^2\equiv
\sum_jc_{u_j}\, \, {\rm ln} (M_{u_j}^2/\mu_u^2)$ for $u=v,t$
(see the appendix).
They are, respectively, the
quartic, quadratic and logarithmic divergences. The divergent terms of the
one loop effective action must be local and covariant, because our
regularization
was studied precisely to preserve covariance. Thus, we can only have
\begin{eqnarray}
i\Gamma_{eff}&=&-{2\over \kappa^2}\int \sqrt{-g} R \, d^4x+
\sum_u a_u \Lambda_u^4\,\int \sqrt{-g} d^4x
\nonumber\\&&+
\sum_u b_u \Lambda_u^2\, \int \sqrt{-g} R\,
d^4x\nonumber\\&&+
\sum_u{\rm ln} \left({\Lambda_u^2\over\mu_u^2}\right)
\,\int \sqrt{-g} (c_uR^2+d_uR_{\mu\nu}
R^{\mu\nu})\, d^4x.
\end{eqnarray}
We have not written the contributions that are finite when the PV
masses tend to infinity.
We do not care about the precise values of the numbers $a_u$, $b_u$, $c_u$
and $d_u$ (in particular, $c_u$ and $d_u$ are well-known \cite{thooft}).
We see that if we want to remove the divergences without modifying the starting
lagrangian, we must set
\begin{eqnarray}
\sum_jc_{u_j}\, M_{u_j}^4\, {\rm ln} {M_{u_j}^2\over\mu_u^2}&=&0,\nonumber\\
\sum_jc_{u_j}\, {\rm ln} {M_{j_u}^2\over\mu_u^2}=0.
\label{logcond3}
\end{eqnarray}
The first condition \cite{anselmi}
avoids the introduction of the cosmological constant, because it was
not present in the starting lagrangian. It is clear that if we start from
a lagrangian
that contains the cosmological term, we do not need
the first condition of (\ref{logcond3}),
because we can reach the same effect by renormalizing the
cosmological constant. The second condition of (\ref{logcond3}) is the
most important, because it is surely dangerous to modify the starting
lagrangian
by adding counterterms quadratic in the Riemann tensor. The resulting
lagrangian
would be renormalizable to all orders but not unitary \cite{stelle}.

The quadratic divergences give the renormalization of the Newton
coupling constant
and the wave function renormalization.
Let
\begin{equation}
\Lambda_u^2=\sum_jc_{u_j}\, M_{u_j}^2\,
{\rm ln} {M_{u_j}^2\over\mu_u^2}\equiv {1\over l_u^2},
\label{length}
\end{equation}
where $l_u$ are lengths that must tend to zero. Let
\begin{eqnarray}
Z&=&1+{1\over 2}\sum_u b_u\left({\kappa \over l_u}\right)^2,\nonumber\\
\kappa^\prime&=&{\kappa\over \sqrt{Z}}.
\label{zek}
\end{eqnarray}
Moreover, set
\begin{equation}
g_{\mu\nu}=\eta_{\mu\nu}+\kappa\phi_{\mu\nu}=\eta_{\mu\nu}+
\kappa^\prime \sqrt {Z}\phi_{\mu\nu}.
\end{equation}
We conclude that if we renormalize the starting lagrangian by replacing it with
\begin{equation}
-2{1\over {\kappa^\prime}^2}R(\eta_{\mu\nu}+\kappa^\prime \sqrt {Z}
\phi_{\mu\nu})=
-2{Z\over \kappa^2}R(\eta_{\mu\nu}+\kappa\phi_{\mu\nu})
\end{equation}
and impose conditions (\ref{logcond3}), then the one loop effective action is
finite. As far as the renormalization is concerned, the quadratic
divergences play the
role that is usually played by the logarithmic ones and
$1\over l_u^2$
play the role of $1\over \varepsilon$. In dimension $r$ the relevant
divergences
are those of degree $r-2$, i.e.\ the highest nonmaximal divergences.

When gravity is coupled to scalar massless field $\varphi$ we have
\begin{eqnarray}
i\Gamma_{eff}&=&\int \sqrt{-g}  \left(-{2\over \kappa^2}R
+{1\over 2}g^{\mu\nu}\partial_\mu\varphi\partial_\nu\varphi
\right)d^4x\nonumber\\&&+
\sum_u\alpha_u\Lambda_u^4\int \sqrt{-g} d^4x+
\sum_u\Lambda_u^2\int \sqrt{-g}
(\beta_u R+\gamma_u \kappa^2 g^{\mu\nu}\partial_\mu\varphi\partial_\nu
\varphi)d^4x
\nonumber\\&&+
\sum_u{\rm ln} {\Lambda_u^2\over\mu_u^2}\int \sqrt{-g}
(\iota_u R^2+\zeta_u R_{\mu\nu}R^{\mu\nu}+\xi_u \kappa^4
(g^{\mu\nu}\partial_\mu\varphi\partial_\nu\varphi)^2
\nonumber\\&&+
\eta_u \kappa^2 R
g^{\mu\nu}\partial_\mu\varphi\partial_\nu\varphi+\nu_u \kappa^2
(g^{\mu\nu}D_\mu D_\nu\varphi)^2)\, d^4x.
\end{eqnarray}
In \cite{thooft} the values of $\iota_u$, $\zeta_u$, $\xi_u$, $\eta_u$ and
$n_u$ can be found ($\ln{\Lambda_u^2\over\mu_u^2}\equiv{2\over r-4}$).
Note that we have not introduced the nonminimal
coupling $\sqrt{-g}R\varphi^2$. Without this term, $\Gamma_{eff}$ turns
out to be very simple as far as the dependence on $\varphi$ is concerned,
because the graphs with external $\varphi$-legs are more convergent.
The coupling $\sqrt{-g}R\varphi^2$ affects this analysis and
$\Gamma_{eff}$ turns out to be nonpolynomial in $\varphi$
(see also \cite{thooft}).

We impose the conditions (\ref{logcond3})
and introduce the definition (\ref{length}) together with
\begin{eqnarray}
Z_g&=&1+{1\over 2}\sum_u\beta_u \left({\kappa\over l_u}\right)^2,\nonumber\\
\kappa^\prime&=&{\kappa\over \sqrt{Z_g}},\nonumber\\
Z_\varphi&=&1-2\sum_u\gamma_u \left({\kappa\over l_u}\right)^2.
\end{eqnarray}
We then substitute the starting lagrangian with
\begin{equation}
\sqrt{-g}\left(-{2\over {\kappa^\prime}^2}R(\eta_{\mu\nu}+
\kappa^\prime \sqrt {Z_g}\phi_{\mu\nu})+{1\over 2}
Z_\varphi g^{\mu\nu}\partial_\mu \varphi
\partial_\nu \varphi\right).
\end{equation}
In this way, the final one loop effective action is finite.

As shown in the previous sections, the second condition of (\ref{logcond3})
depends on the scale $\mu$. The immediate consequence of this fact is that the
finite part of the renormalized one loop effective action depends on $\mu_u$.
However, $\mu_u$ are not related to a coupling constant of the starting
lagrangian and
so they cannot be fixed by normalizing this coupling constant. Thus they
look like physical
constants, that appear only at the level of the effective action, but
not at the level
of the starting lagrangian. They are pure quantum effects, in the
sense that there is no trace of them in the classical lagrangian, but
they appear only in the regularization and renormalization procedures.
They are a sort of quantum coupling constants.
The physical amplitudes surely depend on $\mu_u$
and this permits to measure them, at least in principle.
The variation of $i\Gamma_{eff}$ with respect to
$\mu_u$ is proportional to
\begin{eqnarray}
\int \sqrt{-g}
(\iota_u R^2+\zeta_u R_{\mu\nu}R^{\mu\nu}+\xi_u \kappa^4
(g^{\mu\nu}\partial_\mu\varphi\partial_\nu\varphi)^2
\nonumber\\
+\eta_u \kappa^2 Rg^{\mu\nu}\partial_\mu\varphi\partial_\nu\varphi
+\nu_u\kappa^2(g^{\mu\nu}D_\mu D_\nu\varphi)^2)\, d^4x.
\end{eqnarray}
There are various
different physical amplitudes depending on $\mu_u$.
After having determined the values of
$\mu_u$ (by measuring a scattering between two
gravitons and a scattering between a graviton and
a scalar,
for example), one can plan different experiments (a scattering
between two scalars) to test the values of $\mu_u$.
Note that the arbitrary constants $\mu_u$ are absent
when the space-time
dimension is odd.

\section{Conclusions}
\label{concl}

In the present paper, we studied a technique of the Pauli-Villars type
that regularizes quantum gravity at the one loop order and preserves
explicit covariance. We showed that such a technique cannot work with a
generic gauge-fixing, in particular with the usual one. It surely works with
the gauge-fixing inherited from the formalism of the background field method.
Then we studied the properties of this regularization
technique in connection with the problem of removing the
one loop divergences from the effective action. We showed that the imposition
of some conditions on the PV coefficients avoids the introduction of
dangerous counterterms. We stressed the importance of the quadratic
divergences, because these divergences are responsible for the renormalization
of the Newton coupling constant. Moreover, we showed that two arbitrary
constants appear in the one loop effective action. In four
dimensions, they multiply terms
that are quadratic in the Riemann tensor and could have physical meaning.
They are purely quantum effects.
The renormalization conditions are chosen in order to preserve the form of
the starting lagrangian (Einstein gravity, eventually coupled to matter).
No novelty is brought by our technique in the case of renormalizable theories.
Our reasonings work in any
dimension (greater than two) and in presence of matter. One can hope
that quantum gravity is made finite after treating a certain number of loop
orders (if possible)
in a way similar to the way that we propose for treating
the one loop order. In the most optimistic
case, the one loop order could be enough.

\vspace{24pt}

\begin{center}
{\bf Acknowledgements}
\end{center}

\vspace{12pt}

I am grateful to P. Menotti and F. Vissani for helpful discussions.

\vspace{24pt}

{\bf Appendix. The structure of the one loop
divergences in the PV masses}
\vspace{12pt}

In this appendix we show the standard calculation that one finds
when uses our regularization and employs the background field method.
We restrict to $r=4$, for simplicity.
One has an integral of the form
\begin{equation}
\int{d^4k\over (2\pi)^4}{\sum_j}^\prime c_j \left\{\prod_{i=1}^s{1\over
(k+p_i)^2+M_j^2}\right\}
P_{2s}(k),
\end{equation}
where $p_i$ are linear combinations of the external momenta,
$P_{2s}(k)$ is
a polynomial function of the momenta of degree $2s$ in the momentum $k$
and the sum ${\sum_j}^\prime$ is
from $j=0$ to $j=n$, with the convention $M_0=0$ and $c_0=2$ in the
case of the ghost loops (regulated by the complex vector $W$) and
$c_0=-1$ in the case of graviton loops (regulated by the real tensor $T$).
The integral is convergent due to the regularization conditions
(\ref{pvcond}). One then writes, as usual,
\begin{equation}
\prod_{i=1}^s{1\over
(k+p_i)^2+M_j^2}={\rm const.}
\int_0^1\left(\prod_{i=1}^sdx_i\right)\delta(1-\sum_{i=1}^sx_i)
{1\over ({k^\prime}^2+{M_j^\prime}^2)^s},
\end{equation}
where the constant is not needed to our purposes, while
\begin{eqnarray}
{k^\prime}&=&k+\sum_{i=1}^sx_ip_i,\nonumber\\
{M_j^\prime}^2&=&M_j^2+\sum_{i=1}^sx_ip_i^2-(\sum_{i=1}^sx_ip_i)^2.
\end{eqnarray}
The $x_i$ integration can be brought outside the $k$ integration.
It will be omitted in the following.
The integrand gives a convergent integral even
when the integration variable is taken to be $k^\prime$,
so we can write
\begin{equation}
\int{d^4k\over (2\pi)^4}{\sum_j}^\prime c_j
{P^\prime_{2s}(k)\over (k^2+{M_j^\prime}^2)^s}.
\end{equation}
The polynomial $P^\prime_{2s}(k)$ can be replaced, due to
the symmetries of the integral, by a polynomial $P_s^{\prime\prime}(k^2)$.
We introduce a cut-off $\Lambda$, integrate $|k|$ form $0$ to $\Lambda$
and then take the limit $\Lambda\rightarrow\infty$. The result
is proportional to
\begin{equation}
{\sum_j}^\prime c_j \int_0^\Lambda |k|^3d|k|{P_s^{\prime\prime}(|k|^2)\over
(|k|^2+{M_j^\prime}^2)^s}.
\end{equation}
Let us make the change of variables $y=|k|^2/{M_j^\prime}^2$. We have
\begin{equation}
{\sum_j}^\prime c_j ({M_j^\prime}^2)^2\int_0^{\Lambda^2\over{M_j^\prime}^2}
ydy{1\over
(y+1)^s}\left(\alpha y^s+{\beta(p_i)\over {M_j^\prime}^2}y^{s-1}+
{\gamma(p_i)\over {M_j^\prime}^4}y^{s-2}+\ldots\right).
\end{equation}
Due to (\ref{pvcond}), in the limit $\Lambda
\rightarrow\infty$, we get (apart from some $M_j$-independent
term) a sum of expressions of the form
\begin{equation}
{\sum_j}^\prime c_j({M_j^\prime}^2)^q{\rm ln} {{M_j^\prime}^2\over \mu^2},
\label{appe}
\end{equation}
with $q=0,1,2$,
where $\mu$ is any scale. The divergent parts of these expressions have the
form
\begin{equation}
{\sum_j}c_j({M_j^\prime}^2)^q{\rm ln} {{M_j}^2\over \mu^2}.
\end{equation}
We have thus shown that the one loop divergences can only be of
the form displayed
by Eq. (\ref{logcond2}). (Actually we have only treated the case $r=4$, but the
technique extends to all values of $r$).
We see that the finite part of the result {\sl depends}
on $\mu$ (check the $j=0$ contribution to the sum (\ref{appe})).


\begin{thebibliography}{99}
\bibitem{anselmi} D.\ Anselmi, Phys.\ Rev.\ D 45 (1992) 4473.
\bibitem{anselmi2} D.\ Anselmi, SISSA preprint 18/92/EP , to appear in
Phys.\ Rev.\ D.
\bibitem{olandesi} A.\ Diaz, W.\ Troost, P.\ van Nieuwenhuizen, A.\ van
Proeyen,
Int.\ J.\ Mod.\ Phys.\ A 4 (1989) 3959.
\bibitem{stelle} K.S.\ Stelle, Phys.\ Rev.\ D 16 (1977) 953.
\bibitem{thooft}  G.\ 't Hooft, M.\ J.\ Veltman,
Ann.\  Inst.\  H.\  Poincar\'e 20 79 (1974).
\bibitem{paulivillars}  W.\ Pauli, F.\ Villars, Rev.\  Mod.\  Phys.\
21 434 (1949).
\bibitem{medrano} D.\ M.\ Capper, G.\ Leibbrandt, M.\ Ramon Medrano,
Phys.\ Rev.\  D 12 4320 (1973).
\bibitem{marcus}  L.\ F.\ Abbot, Nucl.\ Phys.\ B185 189 (1981).
\bibitem{sagnotti} M.\ H.\ Goroff, A.\ Sagnotti, Nucl.\ Phys.\ B266
709 (1986).
\bibitem{diagrammar} G.\ 't Hooft, M.\ J.\ Veltman, ``Diagrammar'', CERN
report 73-9, 1973.
\end{thebibliography}
\end{document}